\documentclass{actapoly}
\pdfoutput=1

\usepackage{bm}
\usepackage{xcolor}

\begin{document}

\title[Stability of BECs in a $\mathcal{PT}$ symmetric double-$\delta$
potential]{Stability of Bose-Einstein condensates in a $\mathcal{PT}$ symmetric
  double-$\delta$ potential close to branch points}

\institution{1}{Institut f\"ur Theoretische Physik 1, Universit\"at Stuttgart,
  70550 Stuttgart, Germany}

\author[A. L\"ohle]{Andreas L\"ohle}{1}
\correspondingauthor[H. Cartarius]{Holger Cartarius}{1}
{Holger.Cartarius@itp1.uni-stuttgart.de}
\author[D. Haag]{Daniel Haag}{1}
\author[D. Dast]{Dennis Dast}{1}
\author[J. Main]{J\"org Main}{1}
\author[G. Wunner]{G\"unter Wunner}{1}

\begin{abstract}
  A Bose-Einstein condensate trapped in a double-well potential, where atoms
  are incoupled to one side and extracted from the other, can in the mean-field
  limit be described by the nonlinear Gross-Pitaevskii equation (GPE) with
  a $\mathcal{PT}$ symmetric external potential. If the strength of the
  in- and outcoupling is increased two $\mathcal{PT}$ broken states
  bifurcate from the $\mathcal{PT}$ symmetric ground state. At this bifurcation
  point a stability change of the ground state is expected. However, it is
  observed that this stability change does not occur exactly at the bifurcation
  but at a slightly different strength of the in-/outcoupling effect.
  We investigate a Bose-Einstein condensate in a $\mathcal{PT}$ symmetric
  double-$\delta$ potential and calculate the stationary states. The ground
  state's stability is analysed by means of the Bogoliubov-de Gennes equations
  and it is shown that the difference in the strength of the in-/outcoupling
  between the bifurcation and the stability change can be completely explained
  by the norm-dependency of the nonlinear term in the Gross-Pitaevskii equation.
\end{abstract}


\maketitle

\section{Introduction}

In quantum mechanics non-Hermitian Hamiltonians with imaginary potentials
have become an important tool to describe systems with loss or gain effects 
\cite{Moiseyev2011a}. Non-Hermitian $\mathcal{PT}$ symmetric Hamiltonians,
i.e. Hamiltonians commuting with the combined action of the parity
($\mathcal{P}$: $x \to -x$, $p \to -p$) and time reversal ($\mathcal{T}$:
$x \to x$, $p \to -p$, $\mathrm{i} \to -\mathrm{i}$) operators, possess the
interesting property that, in spite of the gain and loss, they can
exhibit stationary states with real eigenvalues \cite{Bender98}. When the
strength of the gain and loss contributions is increased typically pairs of
these real eigenvalues pass through an exceptional point, i.e.\ a branch
point at which both the eigenvalues and the wave functions are identical, and
become complex and complex conjugate.

A promising candidate for the realisation of a $\mathcal{PT}$ symmetric
quantum system are Bose-Einstein condensates. At sufficiently low temperatures
and densities they can in the mean-field limit be described by the nonlinear
Gross-Pitaevskii equation \cite{Gross61a,Pitaevskii61a}. If a condensate is
trapped in a double-well potential it is possible to add atoms to one side
of the double well and remove atoms from the other. This leads to a gain or
loss to the condensate's probability amplitude. If the strength of both
contributions is equal, the process can effectively be described by a complex
external potential $V(-x) = V^{\ast}(x)$ rendering the Hamiltonian $\mathcal{PT}$
symmetric \cite{Klaiman08a}. The experimental realisation of an open quantum
system with $\mathcal{PT}$ symmetry will be an important step since the
experimental verification has only been achieved in optics so far
\cite{Guo09,Rueter10}.

The nonlinearity $\propto |\psi(\bm{x},t)|^2$ of the Gross-Pitaevskii equation
introduces new interesting properties \cite{Cartarius12b,Cartarius12c,%
  Heiss13a,Dast13a,Dast13b,Cartarius13a,Graefe08a,Graefe08b,Graefe10,Graefe12b}
such as $\mathcal{PT}$ broken states which appear for gain/loss contributions
lower than those at which the corresponding $\mathcal{PT}$ symmetric states
vanish. In optics the same effect can be observed for wave guides with a Kerr
nonlinearity. They are described by an equation being mathematically
equivalent to the Gross-Pitaevskii equation. It has been shown that the
additional features appearing in the presence of the nonlinearity might be
exploited to create uni-directional structures \cite{Ramezani10} or solitons
\cite{Musslimani08a,Abdullaev10,Bludov10,Driben2011a,Abdullaev2011,%
  Bludov2013a}. Since $\mathcal{PT}$ symmetry in optics is extensively studied
\cite{Ramezani10,Musslimani08a,Abdullaev10,Bludov10,Driben2011a,%
  Abdullaev2011,Bludov2013a,Makris08,Makris10,Rausschhaupt2005a,%
  El-Ganainy2007a,Klaiman08a} and has been experimentally
realised \cite{Guo09,Rueter10} these approaches seem to be very promising.

From the mathematical point of view a new type of bifurcation appearing in
the nonlinear $\mathcal{PT}$ symmetric Gross-Pitaevskii equation is of
special interest \cite{Cartarius12b,Cartarius12c,Heiss13a,Dast13a,Dast13b,%
  Cartarius13a,Graefe08a,Graefe08b,Graefe10,Graefe12b}. As in linear quantum
systems two $\mathcal{PT}$ symmetric stationary states merge in an exceptional
point if the strength of a parameter describing the gain and loss processes
is increased. However, in contrast to its linear counterpart the
Gross-Pitaevskii equation possesses no $\mathcal{PT}$ broken states emerging at
this exceptional point. They already appear for lower strengths of the
gain/loss parameter and bifurcate from one of the  $\mathcal{PT}$ symmetric
states in a pitchfork bifurcation. The new bifurcation point has been
identified to be a third-order exceptional point \cite{Heiss13a}. For
attractive nonlinearities one finds that the $\mathcal{PT}$ broken solutions
bifurcate from the ground state. In this scenario the $\mathcal{PT}$ symmetric
ground state is the only state which exists on both sides of the bifurcation
and always possesses a real energy eigenvalue. The pitchfork bifurcation is
expected to entail a change of its stability. However, it is observed that
this stability change does not occur exactly at the bifurcation but at a
slightly different value of the gain/loss parameter.

The discrepancy between the points of bifurcation and stability change seems
to be surprising and does not appear in all similar systems. The mean-field
limit of a two mode approximation with a Bose-Hubbard Hamiltonian 
\cite{Graefe08a,Graefe08b,Graefe10,Graefe12b} does not show this effect.
The model has, however, two crucial differences to the treatment of
Bose-Einstein condensates with the full Gross-Pitaevskii equation in
Ref.\ \cite{Dast13a}. The latter system contains a harmonic trap in which
infinitely many stationary states can be found, whereas the nonlinear
two-mode system exhibits only four states, viz.\ the two
$\mathcal{PT}$ symmetric and the two $\mathcal{PT}$ broken states mentioned
above. Furthermore the nonlinearity derived in \cite{Graefe08a,Graefe08b}
is slightly different from that of the Gross-Pitaevskii equation. It has
the form $\propto |\psi|^2/||\psi||^2$, and hence does not depend on the norm of
the wave function. Thus, there might be two reasons for the appearance of the
discrepancy. It could have its origin in the existence of higher modes
influencing the ground state's stability or in the norm-dependency of the
Gross-Pitaevskii nonlinearity.

It is the purpose of this article to clarify this question. To do so we
study a Bose-Einstein condensate in an idealised double-$\delta$ trap
\cite{Cartarius12b,Cartarius12c,Heiss13a}, a system of which already its
linear counterpart helped to understand basic properties of $\mathcal{PT}$
symmetric structures \cite{Jakubsky05,Mostafazadeh2006a,Mostafazadeh2009a,%
  Mehri-Dehnavi2010a,Jones2008a}. This system is described by the
Gross-Pitaevskii equation, i.e. the contact interaction has the norm-dependent
form $\propto |\psi(\bm{x},t)|^2$. However, it exhibits only four stationary
states of which two are $\mathcal{PT}$ symmetric and two are $\mathcal{PT}$
broken as in the two-mode model \cite{Graefe08a,Graefe08b,Graefe10,Graefe12b}.
Additionally, in a numerical study the structure of the nonlinearity can easily
be changed such that the system's mathematical properties can be brought in
agreement with the mean-field limit of the Bose-Hubbard dimer.

The article is organised as follows. We will introduce and solve the
Gross-Pitaevskii equation of a Bose-Einstein condensate in a double-$\delta$
trap for an attractive atom-atom interaction in Section \ref{sec:BEC_delta}.
Some properties of the stationary solutions which are important for the
following discussions are recapitulated. Then we will investigate the ground
state's stability in the vicinity of the bifurcation in Section
\ref{sec:stability}. The Bogoliubov-de Gennes equations are solved for both
types of the nonlinearity, and the origin of the discrepancy between the
bifurcation and the stability change is discussed. Conclusions are drawn in
Section \ref{sec:conclusion}.

\section{Bose-Einstein condensates in the 
  $\mathcal{PT}$ symmetric double-$\delta$ trap}
\label{sec:BEC_delta}

We assume the condensate to be trapped in an idealised trap of two delta
functions, i.e.\ the potential has the shape \cite{Cartarius12b}
\begin{equation}
  V(x) = -\left( 1- \mathrm{i} \gamma \right) \delta \left(x - b\right) 
  - \left( 1 + \mathrm{i} \gamma \right) \delta \left(x + b\right) \; ,
\end{equation}
where the units are chosen such that the real part due to the action of the
$\delta$ functions has the value -1. It describes two symmetric infinitely thin
potential wells at positions $\pm b$. In the left well we describe an outflux of
atoms by a negative imaginary contribution $-\mathrm{i} \gamma$, and on the
right side an influx of particles is described by a positive imaginary part
$+\mathrm{i} \gamma$ of the same strength. This leads to the time-independent
Gross-Pitaevskii equation in dimensionless form \cite{Cartarius12b},
\begin{multline}
  \biggl [-\frac{\mathrm{d}^2}{\mathrm{d}x^2} -\left( 1- \mathrm{i} \gamma 
    \right) \delta \left(x - b \right) - \left( 1 + \mathrm{i} \gamma \right) 
      \delta \left(x + b\right) \\
      + g |\psi(x) |^2 \biggr ] \psi(x)  = -\kappa^2 {\psi}(x)
      \label{eq:GPE}
\end{multline}
with the energy eigenvalue or chemical potential $\mu = -\kappa^2$. The
parameter $g$ is determined by the s-wave scattering length, which effectively
describes the van der Waals interaction for low temperatures and densities.
Physically it can be tuned via Feshbach resonances. Throughout this article we
assume $g$ to be negative, i.e.\ the atom-atom interaction is attractive.

Solutions to the Gross-Pitaevskii equation are found with a numerical exact
integration. The wave function is integrated outward from $x = 0$ in positive
and negative direction. To do so the initial values $\psi(0), \psi^\prime(0)$,
and $\kappa$ have to be chosen. The arbitrary global phase is exploited such
that $\psi(0)$ is chosen to be real. Together with $\psi^\prime(0) \in
\mathbb{C}$ and $\kappa \in \mathbb{C}$ we have five parameters which have
to be set such that physically relevant wave functions are obtained. These have
to be square-integrable and normalised in the nonlinear system. The five
conditions $\psi(\infty) \to 0$, $\psi(-\infty) \to 0$, and $||\psi|| = 1$
ensure that these conditions are fulfilled, and, together with the five
initial parameters, define a five-dimensional root search problem, which is
solved numerically.

\begin{figure}[tb]
  \centering
  \includegraphics[width=\linewidth]{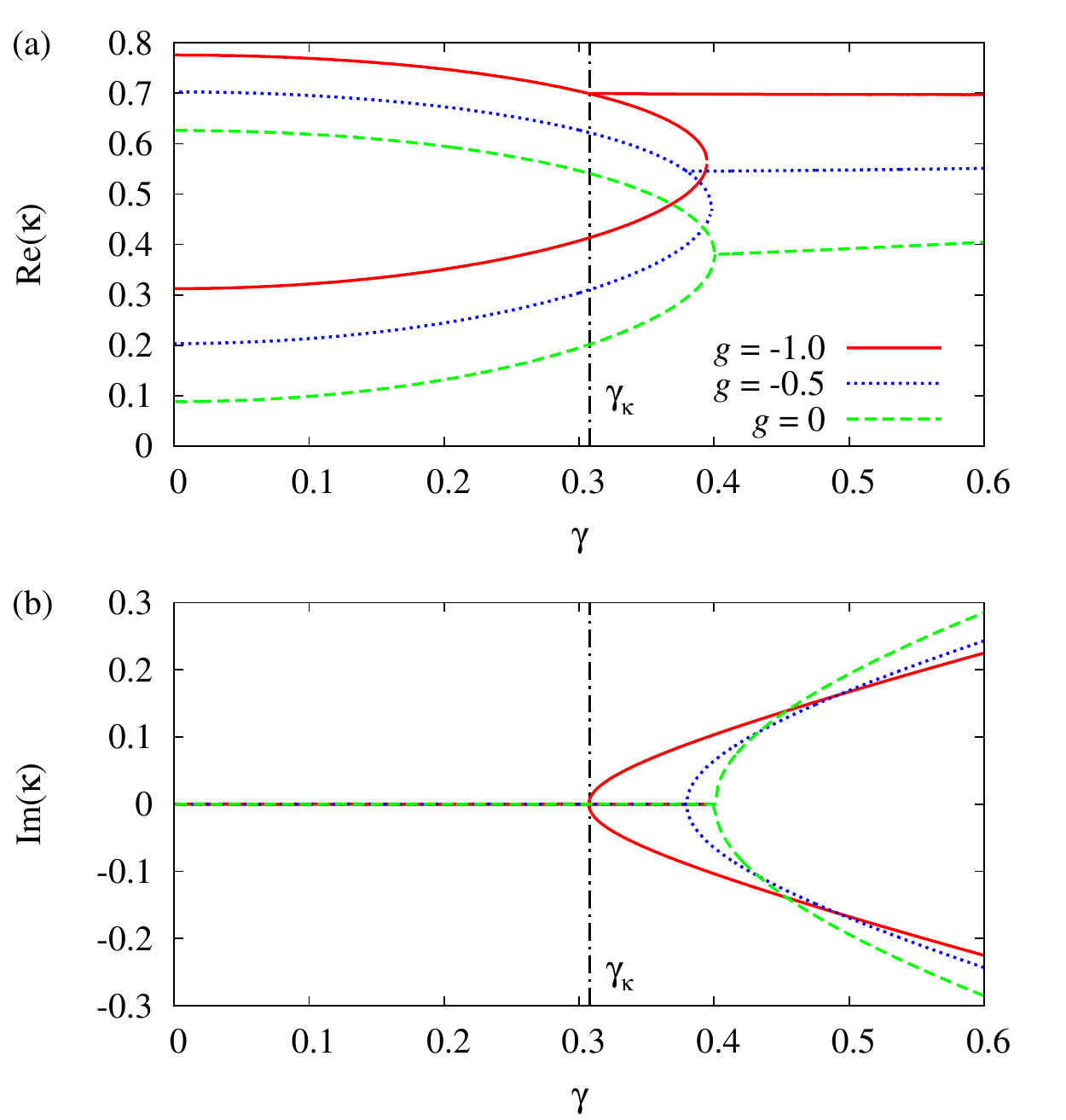}
  \caption{Real (a) and imaginary (b) parts of the stationary solutions found
    for the Gross-Pitaevskii equation \eqref{eq:GPE} for the case $g=-1$
    (red solid lines), which is compared with its linear counterpart $g=0$
    (green dashed lines) and the weaker interaction $g=-0.5$ (blue dotted
    lines). In the nonlinear case we observe two complex conjugate states
    bifurcating from the ground state in a pitchfork bifurcation. For $g=-1$ 
    this occurs at $\gamma_\kappa \approx 0.3071$.}
  \label{fig:stationary}
\end{figure}

Figure \ref{fig:stationary}
shows a typical example for all stationary states found in the case $g=-1$
(red solid lines). Two states with purely real eigenvalues vanish for
increasing $\gamma$ in an exceptional point. Since $\kappa$ is plotted and
$\mu = -\kappa^2$ is the energy eigenvalue, the upper line corresponds to
the ground state. Two complex and complex conjugate eigenvalues bifurcate from
this ground state in a pitchfork bifurcation at a critical value $\gamma_\kappa
\approx 0.3071$. The same plots for $g=-0.5$ (blue dotted lines) and 
$g=0$ (green dashed lines) demonstrate how the pitchfork bifurcation is
introduced by the nonlinearity.

This eigenvalue structure is very generic for $\mathcal{PT}$ symmetric systems
with a quadratic term in the Hamiltonian. It is found for Bose-Einstein
condensates in a true spatially extended double well in one and three
dimensions \cite{Dast13a,Dast13b} which contain the nonlinearity $\propto 
|\psi|^2$ as well as for the mean-field limit of the Bose-Hubbard dimer
\cite{Graefe08a,Graefe08b,Graefe10,Graefe12b} with a term $\propto 
|\psi|^2/||\psi|^2$. The difference in the norm-dependency of the nonlinearity
between both systems does not lead to different eigenvalues, which has already
been mentioned \cite{Graefe12b}. The difference appears, however, for the
dynamics, where the norm plays a crucial role in the presence of gain and
loss \cite{Cartarius13a}.

\section{Stability analysis of the ground state}
\label{sec:stability}

The linear stability is analysed with the Bogoliubov-de Gennes equations. They
are derived under the assumption that a stationary state $\psi_0(x,t)$ is
perturbed by a small fluctuation $\theta(x,t)$, i.e.\
\begin{subequations}
  \begin{equation}
    \psi(x,t) = \mathrm{e}^{\mathrm{i}\kappa^2 t}  \left [ \psi_0(x) 
      + \theta(x,t) \right ] \; ,
    \label{eq:theta}
  \end{equation}
  where
  \begin{equation}
    \theta(x,t) = u(x) \mathrm{e}^{-\mathrm{i}\omega t} + v^*(x)
    \mathrm{e}^{\mathrm{i}\omega^* t} \; .
    \label{eq:ansatz_BdGE}
  \end{equation}
\end{subequations}
With this ansatz and a linearisation in the small quantities $u$ and $v$ one
obtains from the Gross-Pitaevskii equation the coupled system of the
Bogoliubov-de Gennes differential equations, 
\begin{subequations}
  \begin{align}
    \frac{\mathrm{d}^2}{\mathrm{d}x^2} &u(x) = \bigl [-\left( 1 + \mathrm{i}
      \gamma \right) \delta \left(x + b\right) - \left( 1 - \mathrm{i} \gamma
    \right) \delta \left(x - b\right) \notag \\ 
    & + \kappa^2 -\omega + 2 g |\psi_0(x)|^2 \bigr ] u(x) + g \psi_0(x)^2 
    v(x) \; , \\
    \frac{\mathrm{d}^2}{\mathrm{d}x^2} &v(x) = \bigl [ -\left( 1 - \mathrm{i}
      \gamma \right) \delta \left(x + b\right) - \left( 1 + \mathrm{i} 
      \gamma \right) \delta \left(x - b\right) \notag \\
    &+(\kappa^2)^* + \omega + 2 g |\psi_0(x)|^2 \bigr ] v(x) 
    + g \psi_0^*(x)^2 u(x) \; .
  \end{align}
  \label{eq:BdGE}
\end{subequations}

In Equation \eqref{eq:ansatz_BdGE} it can be seen that $\omega$ decides on the
temporal evolution of the fluctuation. Real values of $\omega$ describe
stable oscillations, whereas imaginary parts lead to a growth or decay of
the fluctuation's amplitude. Thus, $\omega$ measures the stability of the
stationary solution $\psi_0$ against small fluctuations.

Numerically the Bogoliubov-de Gennes equations are solved with the same method
as the stationary states, i.e.\ the wave functions $u$ and $v$ are
integrated outward from $x=0$. It can easily be seen that the Bogoliubov-de
Gennes equations \eqref{eq:BdGE} are invariant under the transformation
$u(x) \to u(x)\mathrm{e}^{\mathrm i\chi}$, $v(x) \to v(x)\mathrm{e}^{\mathrm i\chi}$
with a real phase $\chi$. Similarly to the procedure for the integration of
the stationary states this symmetry can be exploited. The remaining initial 
values with which the integration has to be started are $u(0) \in \mathbb{C}$,
$\mathrm{Re}(v(0))$, $u^\prime(0) \in \mathbb{C}$, $v^\prime(0) \in \mathbb{C}$,
$\omega \in \mathbb{C}$. In a nine-dimensional root search they have to be
chosen such that the conditions $u(\pm\infty) \to 0$, $v(\pm\infty) \to 0$,
and
\begin{equation*}
  \int_{-\infty}^{\infty} |u(x)+v^\ast(x)|^2 \,\mathrm{d}x = 1
\end{equation*}
are fulfilled \cite{Dast13a}.

Two further symmetries can be exploited to reduce the number of independent
stability eigenvalues which have to be calculated. The replacement
$(u,v,\omega) \to (v^\ast,u^\ast,-\omega^\ast)$ leaves the ansatz
\eqref{eq:ansatz_BdGE} invariant. Thus, if $\omega$ is a stability eigenvalue
also $-\omega^\ast$ is a valid solution. Furthermore for every eigenvalue
$\omega$ there is also one solution with the eigenvalue $\omega^\ast$ if
the stationary state $\psi_0$ is $\mathcal{PT}$ symmetric. This can be verified
by applying the $\mathcal{PT}$ operator to the Bogoliubov-de Gennes equations
\eqref{eq:BdGE}. Due to these two symmetries it is sufficient to search for
stability eigenvalues with $\mathrm{Re}(\omega) > 0$ and $\mathrm{Im}(\omega)
> 0$.

\subsection{Stability in the vicinity of the bifurcation}

\begin{figure}[tb]
  \centering
  \includegraphics[width=\linewidth]{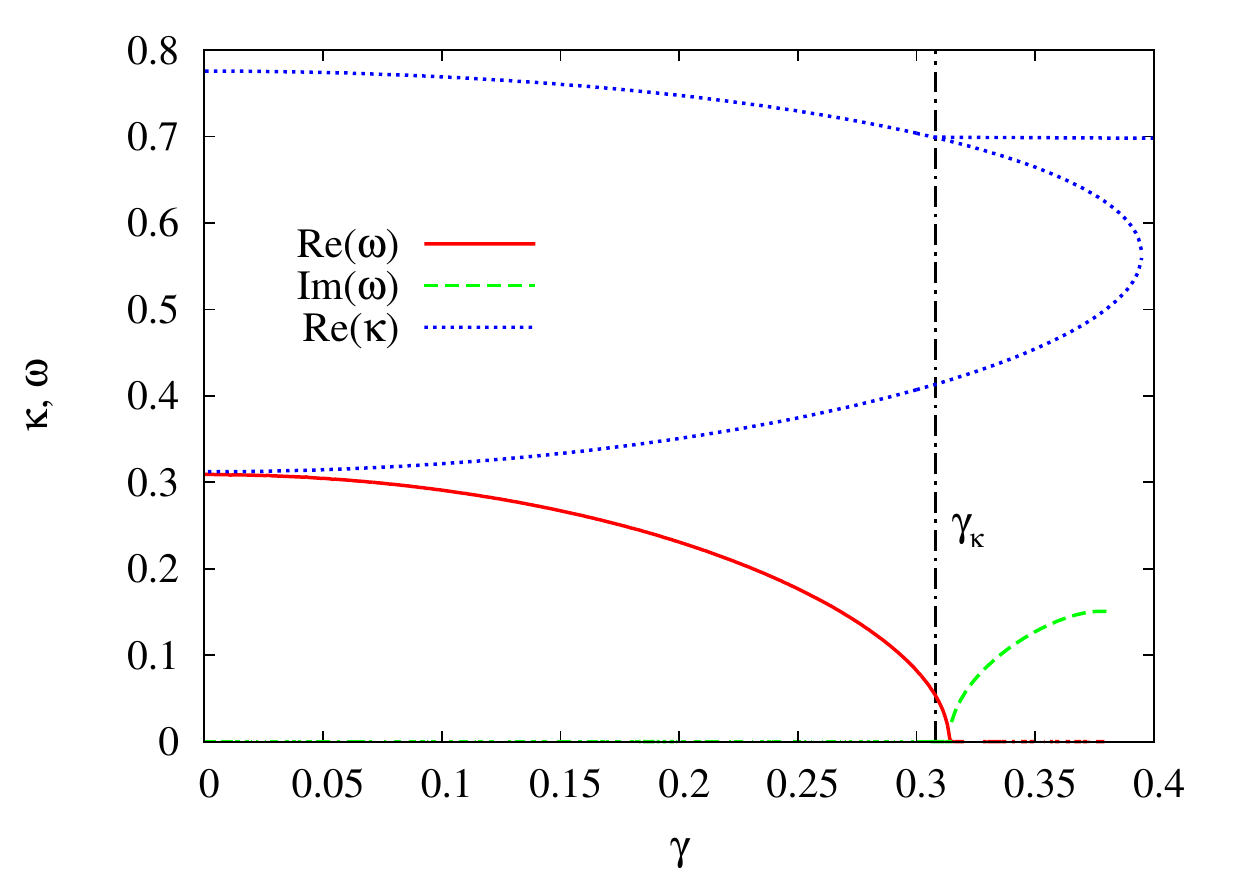}
  \caption{Real (red solid line) and imaginary (green dashed line) part of the
  stability eigenvalue $\omega$ for the stationary ground state in the case
  $g=-1$. The stability change occurs at $\gamma_\omega \approx 0.3138$. To
  illustrate the pitchfork bifurcation the real parts of $\kappa$ of all
  stationary states are also shown (blue dotted lines). The value
  $\gamma_\kappa \approx 0.307$ is marked by the black dashed-dotted line.
  Obviously there is a discrepancy between both values.}
  \label{fig:stability_normdep}
\end{figure}

The relevant question which has to be answered by our calculation is whether
or not the discrepancy between the $\gamma$ values of the pitchfork
bifurcation and the stability change appears for the double-$\delta$ potential.
Thus we calculated the stability eigenvalue with $\mathrm{Re}(\omega) > 0$, 
$\mathrm{Im}(\omega) > 0$ for a range of $\gamma$ around the bifurcation, which
is shown in Figure \ref{fig:stability_normdep}
for $g=-1$. For increasing $\gamma$ the eigenvalue $\omega$ switches from real
to imaginary at $\gamma_\omega \approx 0.3138$ marking the stability change.
The pitchfork bifurcation is visible in the real parts of $\kappa$ of all
stationary states of the system. It is marked by the black dashed-dotted line.
Obviously there is a discrepancy between $\gamma_\kappa$ and $\gamma_\omega$.
The difference
\begin{equation}
  \Delta \gamma = \gamma_\kappa - \gamma_\omega
  \label{eq:difference}
\end{equation}
is $\Delta \gamma \approx -0.0067$.

The system does not possess any further stationary states besides those shown
in Figure \ref{fig:stationary}. Three of these four states are participating
in the pitchfork bifurcation. Only the excited $\mathcal{PT}$ symmetric
solution would be able to influence the dynamics of the ground state at 
a value $\gamma \neq \gamma_\omega$. However, it stays real for all parameters
$\gamma$ shown in Figure \ref{fig:stability_normdep} and cannot cause any
qualitative different behaviour of the ground state's dynamics. Thus, an
influence of further states can be ruled out to be the reason for the
discrepancy in the double-well system of Refs.\ \cite{Dast13a,Dast13b}.

\begin{figure}[tb]
  \centering
  \includegraphics[width=\linewidth]{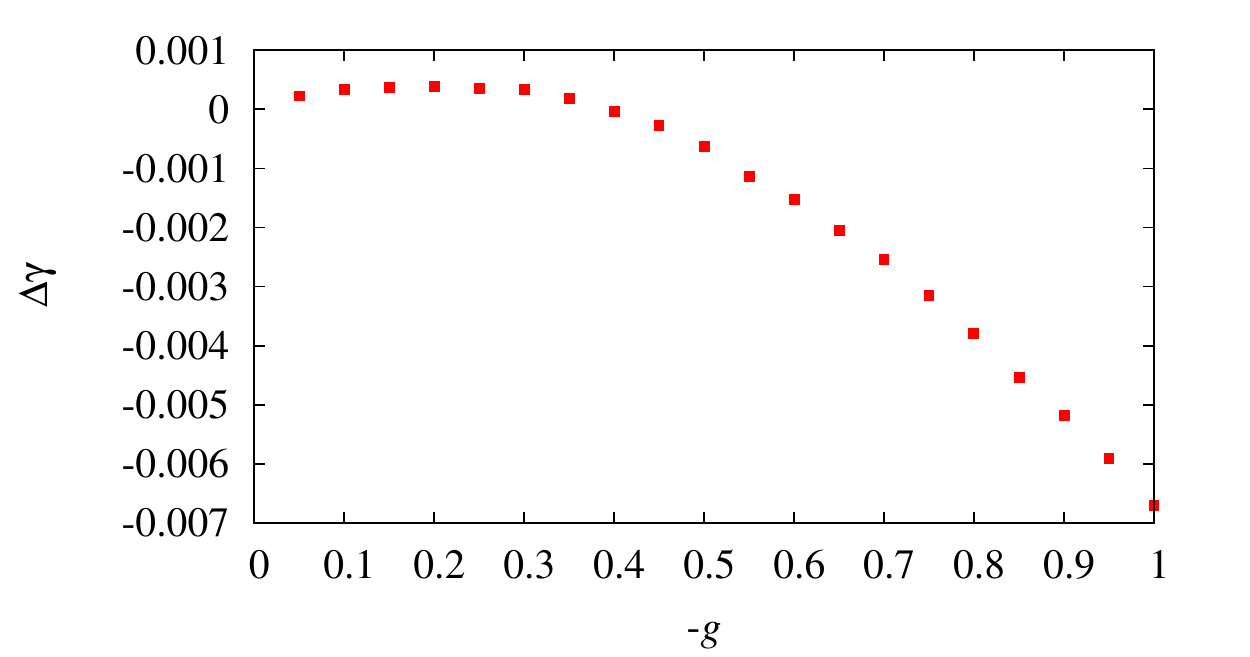}
  \caption{Difference in between $\gamma_\kappa$ and $\gamma_\omega$ as defined
  in Equation \eqref{eq:difference} as a function of $g$. A strong dependency is
  clearly visible.}
  \label{fig:gap_g}
\end{figure}

The remaining difference between the Gross-Pitaevskii equation \eqref{eq:GPE}
and the two mode system of Refs.\ \cite{Graefe08a,Graefe08b,Graefe10,Graefe12b}
is the norm-dependency of the nonlinearity. Indeed, an influence of the norm
is already present if the study done in Figure \ref{fig:stability_normdep} is
repeated for different values of the nonlinearity parameter $g$. Figure
\ref{fig:gap_g}
shows $\Delta \gamma$ as a function of $g$. A strong dependency is visible. Even
the sign changes. For $g \lessapprox 0.4$ the ground state becomes unstable
at $\gamma_\omega < \gamma_\kappa$. For $g\to 0$ the discrepancy vanishes as
expected.

\subsection{Norm-independent variant of the Gross-Pitaevskii
  equation}

An even clearer identification of $\Delta \gamma$ with the norm-dependency of
the nonlinearity in the Gross-Pitaevskii equation \eqref{eq:GPE} can be given
with a small modification. The replacement
\begin{equation}
  g |\psi|^2 \to \frac{g |\psi|^2}{\int |\psi|^2 dx}
\end{equation}
makes the Gross-Pitaevskii equation \eqref{eq:GPE} norm-independent. Note that
this is exactly the form of the mean-field limit of Refs.\ 
\cite{Graefe08a,Graefe08b,Graefe10,Graefe12b}. Since the stationary states
are normalised to 1 they are not influenced by the replacement. However,
it influences the dynamics and also the linear stability. The Bogoliubov-de
Gennes equations have to be adapted. Assuming again a small perturbation
of the form \eqref{eq:theta} and \eqref{eq:ansatz_BdGE} a linearisation in
$u$ and $v$ leads us to
\begin{subequations}
\begin{multline}
  \frac{\mathrm{d}^2}{\mathrm{d}x^2} u(x) = \bigl [ -\left( 1 + \mathrm{i}
    \gamma \right) \delta \left(x + b\right) - \left( 1 - \mathrm{i} \gamma
  \right) \delta \left(x - b\right) \\ 
  +\kappa^2 - \omega + 2 g |\psi_0(x)|^2 \bigr ] u(x) + g \psi_0(x)^2 v(x) \\
  + g |\psi_0(x)|^2 \psi_0(x) S \; ,
  \label{eq:mod_BdGE_u}
\end{multline}
\begin{multline}
  \frac{\mathrm{d}^2}{\mathrm{d}x^2} v(x) = \bigl [ -\left( 1 - \mathrm{i}
      \gamma \right) \delta \left(x + b\right) - \left( 1 + \mathrm{i} 
      \gamma \right) \delta \left(x - b\right) \\ + (\kappa^2)^* + \omega 
    + 2 g |\psi_0(x)|^2 \bigr ] v(x) + g \psi_0^\ast(x)^2 u(x) \\ 
  + g |\psi_0(x)|^2 \psi_0^*(x) S
  \label{eq:mod_BdGE_v}
\end{multline}
with the integral
\begin{equation}
  S = \int \left[v(x) \psi_0(x) + u(x) \psi_0^*(x) \right] dx  \; .
  \label{eq:mod_BdGE_int}
\end{equation}
\end{subequations}

For a numerical solution of the modified Bogoliubov-de Gennes equations the
value of $S$ is included in the root search. i.e.\ for the integration of
the Equations \eqref{eq:mod_BdGE_u} and \eqref{eq:mod_BdGE_int} a value for
$S$ is guessed and subsequently compared with the result of Equation
\eqref{eq:mod_BdGE_int}. Since $S$ is in general a complex value this increases
the dimension of the root search to 11.

\begin{figure}[tb]
  \centering
  \includegraphics[width=\linewidth]{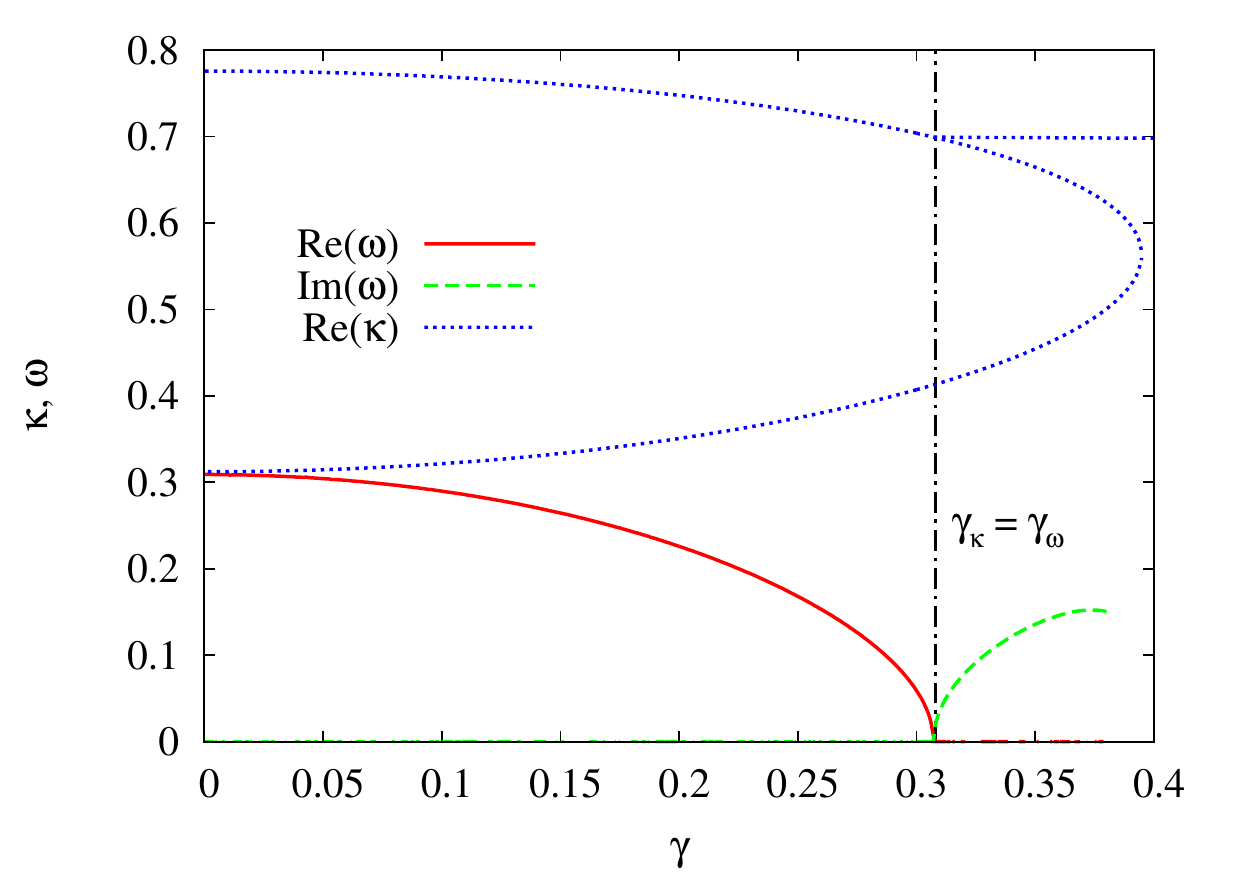}
  \caption{The same as in Figure \ref{fig:stability_normdep} but for the
    modified Bogoliubov-de Gennes equations
    \eqref{eq:mod_BdGE_u}-\eqref{eq:mod_BdGE_int}. Now the $\gamma$ values
    at which the pitchfork bifurcation and the stability change occur
    agree perfectly.}
  \label{fig:stability_normindep}
\end{figure}

An example for a typical result is shown in Figure
\ref{fig:stability_normindep}.
The discrepancy between the $\gamma$ values at which the pitchfork bifurcation
and the stability change occur vanishes. Both values agree as it is expected
for a pitchfork bifurcation. This is true for all values of $g$. Thus, we
conclude that the discrepancy appearing in Figure \ref{fig:stability_normdep}
for the Gross-Pitaevskii equation is solely a result of the norm-dependent
nonlinearity $\propto |\psi|^2$ in the Hamiltonian and is not a consequence of
the interaction with higher excited states.

\subsection{Discussion}

The reason why the stability change does not occur exactly at the bifurcation
can also be understood intuitively. Figure \ref{fig:stationary} indicates that
the value $\gamma_\kappa$ of the pitchfork bifurcation is not equal for all
values of $g$. For $g = 0$ there is no pitchfork bifurcation. The two real
eigenvalues $\kappa$ vanish in a tangent bifurcation and two complex
eigenvalues emerge. Only for nonvanishing $g$ the pitchfork bifurcation exists
and moves to smaller values of $\gamma$ for increasing $|g|$.

If now the norm of a wave function changes, this can also be understood
as a variation of $g$, i.e.\ a wave function with a norm $N=||\psi||^2$
has the same effect as a wave function with the norm 1 and a modified
value $\tilde{g} = Ng$. The values $\gamma_\kappa$ of the pitchfork bifurcation
is obviously different for $g$ and $\tilde{g}$. With this relation in mind it
is not surprising that a fluctuation changing the norm of the wave function
may cause a qualitative change of the condensate's stability properties in the
vicinity of $\gamma_\kappa$.

\section{Conclusions}
\label{sec:conclusion}

We investigated the origin of the discrepancy between the value of the
gain/loss parameters, at which the ground state of a Bose-Einstein condensate
in a double-well trap passes through a pitchfork bifurcation and at which
its stability changes. In a naive expectation these $\gamma$ values should
be identical. However, it is found that this is not exactly fulfilled. Since
this discrepancy does not occur for a similar system, viz.\ the mean-field
limit of a Bose-Hubbard dimer \cite{Graefe08a,Graefe08b,Graefe10,Graefe12b},
we investigated the differences between the equations describing both systems.
It was found that the norm-dependency of the nonlinearity in the Hamiltonian
$\propto |\psi|^2$ is unambiguously the origin of the discrepancy. It can be
completely removed with the replacement $|\psi|^2 \to |\psi|^2/||\psi||^2$. An
intuitive explanation can be given. Fluctuations which change the norm of a
stationary state are able to shift the position $\gamma_\kappa$ of the
bifurcation.

Since the dynamical properties of the wave functions are crucial for an
experimental observability of a $\mathcal{PT}$ symmetric Bose-Einstein
condensate it is important to know about all processes introducing possible
instabilities. As has been shown in this article the stability relations
are nontrivial close to branch points in condensate setups with gain and
loss. The Gross-Pitaevskii has a norm-dependent nonlinearity, and therefore
it should be clarified in future work, which type of fluctuation influences
the wave function's norm such that additional instabilities appear. In
particular, it would be interesting to see, how the amplitude of
a fluctuation is related to the size of the difference $\Delta \gamma$. Also
a deeper understanding, how this effect influences realistic setups generating
the $\mathcal{PT}$ symmetric external potential \cite{Kreibich13a}, would be
of high value.

\begin{acknowledgements}
  We thank Eva-Maria Graefe for stimulating discussions.
\end{acknowledgements}


\end{document}